\documentclass[a4paper,fleqn,usenatbib]{mnras}

\usepackage{newtxtext,newtxmath}
\usepackage[T1]{fontenc}
\usepackage{ae,aecompl}

\usepackage{graphicx}	% Including figure files
\usepackage{amsmath}	% Advanced maths commands
\usepackage{amssymb}	% Extra maths symbols
\usepackage{booktabs}   % enable to use /toprule, /midrule, /bottomrule in tables
\usepackage{multirow}   % ebabke to use multiple rows in tables.
\usepackage{rotating}   % 
\usepackage{natbib}
\usepackage{color, soul} % color and highlight text, e.g., \textcolor{red}{\hl{foo}}
\soulregister\cite7
\soulregister\citet7
\soulregister\citep7
\soulregister\ref7
\usepackage{xcolor} %color text
\usepackage{xspace}
\usepackage{subcaption}     % plot sub figures
\usepackage[stable]{footmisc}
\usepackage{etoolbox} %%fix possible crossref errors
\makeatletter
\patchcmd\@combinedblfloats{\box\@outputbox}{\unvbox\@outputbox}{}{%
	\errmessage{\noexpand\@combinedblfloats could not be patched}%
}%
\makeatother
\title[On the interior composition and structure of terrestrial exoplanets]{Enhanced constraints on the interior composition and structure of terrestrial exoplanets}

\author[H. S. Wang, et al.]{
	H. S. Wang,$^{1,2,\thanks{E-mail: haiyang.wang@anu.edu.au}}$
    F. Liu,$^{3}$
    T. R. Ireland,$^{2,4}$
    R. Brasser,$^{5}$
    D. Yong$^{1}$
    and C. H. Lineweaver$^{1,2}$
	\\
	$^{1}$Research School of Astronomy and Astrophysics, The Australian National University, Canberra, ACT 2611, Australia\\
	$^{2}$Planetary Science Institute, The Australian National University, Canberra, ACT 2611, Australia\\
	$^{3}$Lund Observatory, Department of Astronomy and Astrophysics, Lund University, Box 43, SE-22100 Lund, Sweden\\
	$^{4}$Research School of Earth Sciences, The Australian National University, Canberra, ACT 2601, Australia\\
	$^{5}$Earth Life Science Institute, Tokyo Institute of Technology, Tokyo, 152-8550, Japan
}
\date{Accepted 2018 October 9. Received 2018 October 9; in original form 2018 August 7}

\pubyear{2018}

\begin{document}
\label{firstpage}
\pagerange{\pageref{firstpage}--\pageref{lastpage}}
\maketitle

\begin{abstract}
Exoplanet interior modelling usually makes the assumption that the elemental abundances of a planet are identical to those of its host star. Host stellar abundances are good proxies of planetary abundances, but only for refractory elements. This is particularly true for terrestrial planets, as evidenced by the relative differences in bulk chemical composition between the Sun and the Earth and other inner solar system bodies. The elemental abundances of a planet host star must therefore be devolatilised in order to correctly represent the bulk chemical composition of its terrestrial planets. Furthermore, nickel and light elements make an important contribution alongside iron to the core of terrestrial planets. We therefore adopt an extended chemical network of the core, constrained by an Fe/Ni ratio of 18 $\pm$ 4 (by number). By applying these constraints to the Sun, our modelling reproduces the composition of the mantle and core, as well as the core mass fraction of the Earth. We also apply our modelling to four exoplanet host stars with precisely measured elemental abundances: Kepler-10, Kepler-20, Kepler-21 and Kepler-100. If these stars would also host terrestrial planets in their habitable zone, we find that such planets orbiting Kepler-21 would be the most Earth-like, while those orbiting Kepler-10 would be the least. To assess the similarity of a rocky exoplanet to the Earth in terms of interior composition and structure, high-precision host stellar abundances are critical. Our modelling implies that abundance uncertainties should be better than $\sim$ 0.04 dex for such an assessment to be made.

\end{abstract}


%%%%%%%%%%%%%%%%%%%%%%%%%%%%%%%%%%%%%%%%%%%%%%%%%%

%%%%%%%%%%%%%%%%% BODY OF PAPER %%%%%%%%%%%%%%%%%%

\section{Introduction}
\label{sec:intro}
 We are at the cusp of the golden era of discovery and characterization of exoplanets. To date, more than 3700 exoplanets have been confirmed\footnote{NASA Exoplanet Archive, \url{https://exoplanetarchive.ipac.caltech.edu}, which is operated by the California Institute of Technology, under contract with the National Aeronautics and Space Administration (NASA) 2under the Exoplanet Exploration Program.}
, via a wide range of observational techniques including transit photometry, radial velocity measurement, imaging, and microlensing \citep{Winn2015}. 
Continued improvements in observational and modeling techniques have led to more precise inference of planetary mass and radius \citep{Weiss2016, Stassun2017a, Stassun2017b}.
High-precision and homogeneous analyses of host stellar abundances are also increasingly available \citep[e.g.][]{Nissen2010, Adibekyan2012, DaSilva2015, Adibekyan2015, Brewer2016, Spina2016, Delgado2017, Kos2018}. 
At the same time, our knowledge of our own planet Earth and the Solar System has been enormously expanded with the efforts from the broad communities in geosciences, cosmochemistry, planet formation and star formation 
\citep[e.g.][]{Ireland2000, McDonough2014, Wang2016, Brasser2016, Brasser2016a, Kwok2016, Norris2017, Wang2018}. 
Together these have opened the door for
detailed studies of chemical composition, interior structure and habitability of rocky exoplanets. 

Previous studies concerned with exoplanet interiors have generally assumed differentiated structures and molecular compositions to compute mass-radius relations \citep[e.g.][]{Valencia2007, Seager2007, Zeng2013, Howard2013, Dressing2015}.
It has been concluded that with only mass and radius measurements an exact interior composition cannot be inferred for an exoplanet because the problem is highly underconstrained or degenerate \citep{Rogers2010}. 
With the increasing availability of elemental abundances of planet host stars, recent studies \citep{Dorn2015, Santos2015} have discussed the reduction of degeneracies in constraining the interior composition and structure of rocky exoplanets by adding host stellar abundances as another principal constraint.  
Since then, an increasing number of updated models (with the similar principal constraints) have been proposed \citep[e.g.][]{Unterborn2016, Dorn2017, Dorn2017b, Brugger2017, Unterborn2018}, to improve the modelling of exoplanetary interiors.

However, a major cause of modelling inaccuracy inherent in the prevalent exoplanetary interior models \citep[e.g.][]{Santos2015, Dorn2017, Brugger2017, Unterborn2018, Hinkel2018} is that the elemental abundances of a rocky exoplanet are simplified to be identical to the elemental abundances of its host star. Host stellar abundances are good proxies of planetary abundances, but \textit{only} for refractory elements\footnote{Elements with relatively high equilibrium condensation temperatures ($\gtrsim$ 1360 K) and they are resistant to heat/irradiation/impact.}. This is particularly true for terrestrial planets, as evidenced by the relative
differences in the bulk composition between the Sun, Earth and other inner solar system bodies \citep{Davis2006, Carlson2014}. This led {\citet{Dorn2017b} to conclude that further studies of solar system bodies are needed to improve our understanding of the correlation of relative bulk abundances between planets and host star and of the effect of such abundance correlations on exploring exoplanet interiors.} \cite{Wang2018b} has quantified the devolatilisation (i.e. depletion of volatiles) in going from the solar nebula to the Earth. 
The elemental abundance differences between the Earth and the Sun for Si, Mg, Fe and Ni are slight, but significant (a devolatilisation factor of 10-20\%); those for O, S, and C are substantial and devolatilised by up to 3 orders of magnitude. The former differences, in combination with the substantially devolatilised O, will have a direct and nontrivial effect on the mantle and core composition; the latter differences will have profound impact on the atmospheric composition, including importantly the abundance of surface water, and therefore habitability in general. 

An additional source of modelling inaccuracy in the prevalent studies of exoplanetary interiors is that elements lighter than Fe and Ni are rarely taken into account in the core composition.
Light elements play a key role in compensating the 5-10\% density deficit of Earth's core \citep{Birch1964, Hirose2013, McDonough2014, Wang2018} (compared to pure iron at core pressures) and in differentiating the liquid outer core from the solid inner core. The importance of light elements in a terrestrial exoplanet's core cannot be ignored, and their presence has direct consequences for estimates of core mass fraction and the melting temperature of an outer core (if it exists), and therefore the generation of exoplanetary magnetic fields.

With the goal of reducing the uncertainty in the estimates of interior composition and structure of terrestrial exoplanets, we present several important constraints and our analytical method in Sect. \ref{sec:constraints}. Our modelling results are reported in Sect. \ref{sec:results}, followed by the discussion in Sect. \ref{sec:disc}. We conclude in Sect. \ref{sec:conclusion}.

\section{Constraints and analysis}
\label{sec:constraints}
\begin{table*}
	\centering
	\caption{Predicted bulk elemental composition [X/Al]$^a$ of potential habitable-zone terrestrial exoplanets as devolatilised from their host stellar abundances$^b$.}
	\begin{tabular}{l cc cc ccc} % four columns, alignment for each
		\toprule
		\multirow{1}{*}{X$^c$} & \multirow{1}{*}{$F_D$ (\%) $^d$} & \multirow{1}{*}{$F_D$ (dex) $^d$} & Kepler 10 (L16) & Kepler 10 (S15)$^e$ & Kepler 20 & Kepler 21$^f$ & Kepler 100 \\
		\hline
		
		C & $99.6\pm0.1$& 2.42 $\pm$ 0.09 & -2.42 $\pm$ 0.09 & $-2.29\pm0.10$& -2.48 $\pm$ 0.10 & -2.38 $\pm$ 0.10 & -2.47 $\pm$ 0.12\\
		
		S & $93.4\pm0.6$& $1.18\pm0.04$ & -1.20 $\pm$ 0.04 & \textendash& \textendash & -1.25 $\pm$ 0.06 & -1.23 $\pm$ 0.07\\	
		
		O & $82\pm1$& $0.74\pm0.02$ & -0.67 $\pm$ 0.03 & $-0.39\pm0.08$& -0.86 $\pm$ 0.08 & -0.73 $\pm$ 0.07 & -0.71 $\pm$ 0.10\\		
		
		Na & $75\pm1$& 0.60 $\pm$ 0.02& -0.72 $\pm$ 0.02 & \textendash& -0.65 $\pm$ 0.04 & -0.62 $\pm$ 0.05 & -0.57 $\pm$ 0.04\\	
		
		Si & $20\pm3$& 0.10 $\pm$ 0.02& -0.17 $\pm$ 0.02 & $-0.02\pm0.03$& -0.14 $\pm$ 0.03 & -0.09 $\pm$ 0.03 & -0.13 $\pm$ 0.04\\
		
		Fe & $14\pm3$& 0.07 $\pm$ 0.02& -0.20 $\pm$ 0.02 & $-0.07\pm0.03$& -0.09 $\pm$ 0.06 & -0.09 $\pm$ 0.07 & -0.17 $\pm$ 0.08\\
		
		Mg & $14\pm3$& 0.07 $\pm$ 0.02&-0.10 $\pm$ 0.02 & 0.09 $\pm$ 0.06& -0.05 $\pm$ 0.04 & -0.07 $\pm$ 0.04 & -0.10 $\pm$ 0.05\\	
		
		Ni & 10 $\pm$ 4& 0.05 $\pm$ 0.02 & -0.20 $\pm$ 0.02 & \textendash& -0.07 $\pm$ 0.03 & -0.12 $\pm$ 0.03 & -0.11 $\pm$ 0.03\\	
		
		Ca & $0$& $0$ & -0.05 $\pm$ 0.01 & \textendash& 0.00 $\pm$ 0.04 & 0.00 $\pm$ 0.04 & -0.06 $\pm$ 0.05\\
		
		Al & $0$& $0$ & 0.00 $\pm$ 0.01 & \textendash& 0.00 $\pm$ 0.02 & \textendash & 0.00 $\pm$ 0.03\\

		\bottomrule
		\multicolumn{8}{p{14cm}}{\footnotesize$^a$ $\textrm{[X/Al]} = \log((\textrm{X/Al})_{\textrm{planet}} / (\textrm{X/Al})_{\textrm{Sun}})$, where (X/Al) is the abundance ratio (by number in linear) of an element X to Al, and solar abundance refers to \cite{Asplund2009}. Normalising to a refractory element other than Al does not change our results. Following \cite{Wang2018b}, we choose Al as the normalisation element since it is the most refractory major element.} \\
		
		\multicolumn{8}{p{14cm}}{\footnotesize$^b$ Sources of host stellar abundances: Kepler 10: L16- \citep[Column 2 of Table 2 of][]{Liu2016}; S15- \citep[Row 2 of Table A.1 and Row 2 of Table A.2 in][]{Santos2015}. Kepler 20, Kepler 21 and Kepler 100: Table 3 of \cite{Schuler2015}.} \\	
		
		\multicolumn{8}{p{14cm}}{$^c$ Elements (X) are listed in order of decreasing devolatilisation factor.}\\ 
		
		\multicolumn{8}{p{14cm}}{$^d$ $F_D$: Devolatilisation factor, which refers to Figure \ref{fig:model}. 
		$F(D)$ (\%) of each element is derived by $1 - (\textrm{X/Al})_{\textrm{planet}} / (\textrm{X/Al})_{\textrm{Sun}}$, where $(\textrm{X/Al})_{\textrm{planet}} / (\textrm{X/Al})_{\textrm{Sun}}$ can be read from the left-hand y-axis according to the devolatilisation pattern. The corresponding $F(D)$ (dex) is directly opposite to $\textrm{[X/Al]}$ (= $\log((\textrm{X/Al})_{\textrm{planet}} / (\textrm{X/Al})_{\textrm{Sun}})$), which can be read from the right-hand y-axis according to the devolatilisation pattern.}\\
		
		\multicolumn{8}{p{14cm}}{\footnotesize$^e$ The elemental abundances of Kepler-10 (S15) planets are normalized to Fe abundance in the host star, as no available Al abundance for Kepler-10 is reported in \citet{Santos2015}. This normalisation difference will not change the subsequent analyses of key elemental ratios including Mg/Si, Fe/Si, and C/O.}\\
		
		\multicolumn{8}{p{14cm}}{\footnotesize$^f$ The elemental abundances of Kepler-21 planets are normalized to Ca abundance in the host star, as no available Al abundance for Kepler 21 is documented in \citet{Schuler2015}.}\\		    
	\end{tabular}
	\label{tab:abu}
\end{table*}

\begin{figure*}
	\centering 
	\includegraphics[trim=1.0cm 0.8cm 0.8cm 2.0cm, scale=0.5,angle=90]{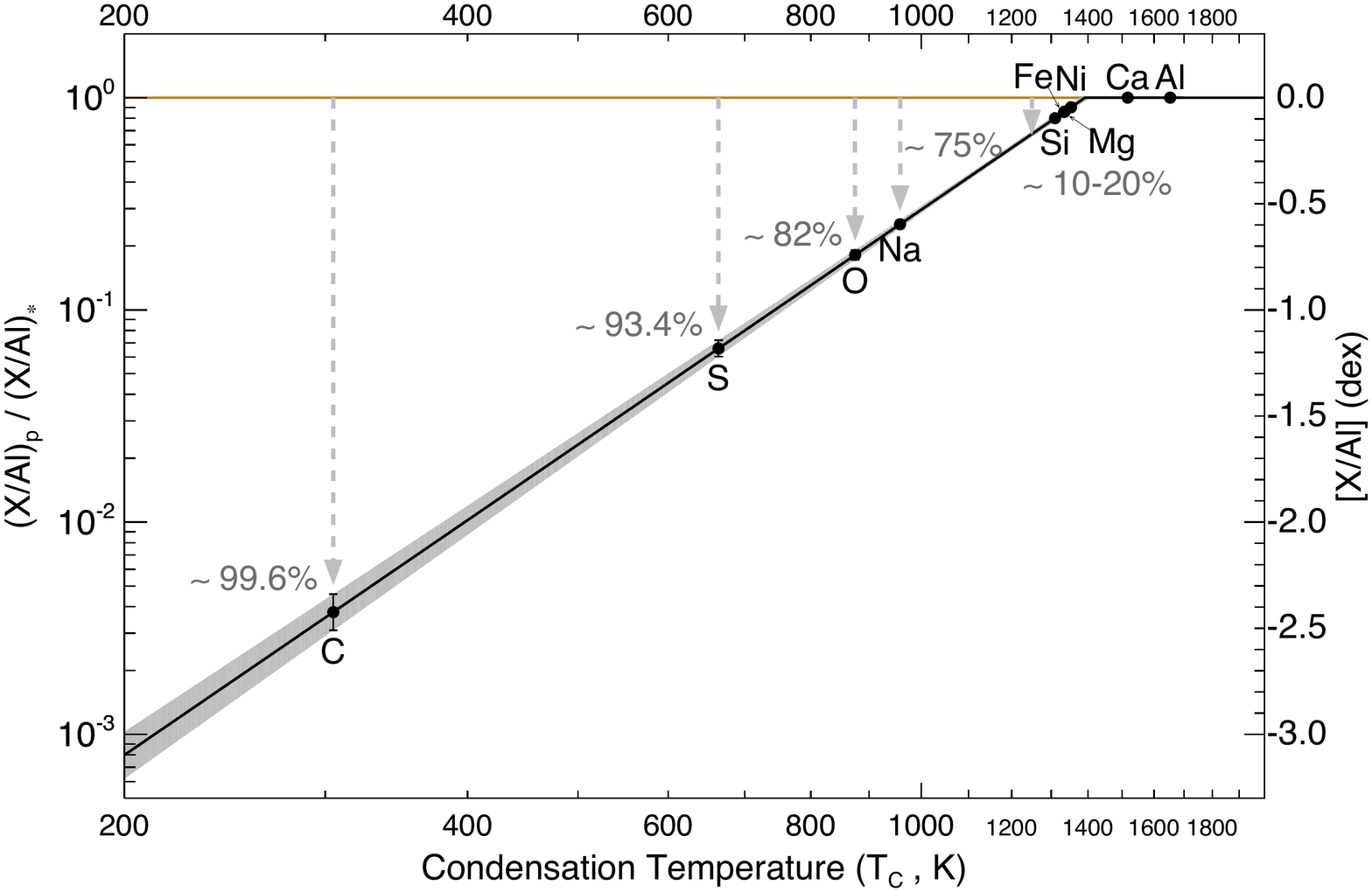}
	\caption{Devolatilisation patterns from stellar nebulae to terrestrial exoplanets, which are drawn from the Sun-to-Earth volatility trend (devolatilisation pattern) of \citet{Wang2018b}, including the best fit (the solid line in black) and its 1$\sigma$ uncertainty (the wedge in grey), as well as the devolatilization factors indicated by the dashed arrows.
	Host stellar abundance is normalized to 1 in linear (or 0 in dex), shown as the solid horizontal line in brown. The x-axis is the elemental 50\% condensation temperature ($T_\textrm{C}$) in logarithm from \citep{Lodders2003}, except for $T_\textrm{C}$(C) and $T_\textrm{C}$(O), which are adopted from \citet{Wang2018b} (see text). The left-hand y-axis is in the logarithmic scale for the planet (\textquoteleft p')-to-star (\textquoteleft $\ast$') elemental abundance ratio, ($\textrm{X/Al})_{\textrm{p}} / (\textrm{X/Al})_{\textrm{*}})$, while the right-hand y-axis is labeled in dex for the corresponding $\textrm{[X/Al]}$ (= $\log((\textrm{X/Al})_{\textrm{p}} / (\textrm{X/Al})_{\textrm{*}})$). Only elements that are essential to control the mineralogy of a terrestrial planet are indicated on this plot and then employed in our subsequent analysis.}
	\label{fig:model}
\end{figure*}

\subsection{Bulk elemental composition of a terrestrial planet}
\label{sec:bulk}

As mentioned earlier, elemental abundances between a terrestrial planet and its host star are not identical, especially for non-refractory or volatile elements. 
We argue that the elemental abundances of the host star of a terrestrial planet should be devolatilised in order to infer the planetary bulk composition. \cite{Wang2018b} has established a devolatilisation fiducial model (see Figure \ref{fig:model}) by quantifying the compositional differences between the proto-Sun and the bulk Earth as a function of elemental condensation temperatures.
This work is built upon the premise that devolatilisation in a nebula may be universal and the Sun-to-Earth volatility trend (devolatilisation pattern) of \cite{Wang2018b} is applicable to infer the bulk composition of terrestrial exoplanets, particularly those within circumstellar habitable zones. Some caveats and limitations of this assumption are discussed in the following paragraph (see Section \ref{sec:limits} for more details).

A variety of outcomes for the bulk composition of a rocky planet may result from composition-, location-, and timescale-dependent differences in various (and sometimes contrary) fractionation processes, such as incomplete condensation \citep[e.g.][]{Wasson1974, Grossman1974}, partial evaporation \citep[e.g.][]{Anders1964, Alexander2001, Braukmuller2018}, accretionary collision \citep[e.g.][]{ONeill2008, Visscher2013, Hin2017}, and giant impact in conjunction with magmatic solidification \citep[e.g.][]{Chen2013, Norris2017, Dhaliwal2018}. The compositional differences of Earth, Mars and Venus should be a measure of these variations within our own Solar System, but the extent to which the bulk compositions of Venus and Mars are different from the Earth is still debated in both the geochemical/cosmochemical community \citep[e.g.][]{Morgan1980, Wanke1988, Taylor2013, Wang2018} and the planet formation community \citep[e.g.][]{Kaib2015, Fitoussi2016, Brasser2017, Brasser2018}. We also note that the \textit{bulk} compositions of Venus and Mars are not well determined yet and thus reliable quantitative analysis of bulk compositional differences between them and the Earth is difficult. We have been conservative to preclude the application of this algorithm to Mercury-orbit-like planets (e.g. warm super-Earths), as such planets may be more devolatilised than predicted due to plausibly more intensive stellar irradiation and bombardment histories.
In spite of the complexity of planet formation, an essential step to improve the study of exoplanetary chemistry is to take into account devolatilization, with the best constrainable trend from the Sun to the Earth.

In Figure \ref{fig:model}, we plot 10 major elements (C, S, O, Na, Si, Fe, Mg, Ni, Ca, and Al) that are essential for estimating the mineralogy of a terrestrial planet and place these on the best-fitting devolatilization pattern of \cite{Wang2018b}. These 10 elements account for more than 99\% of the total mass of Earth \citep{McDonough2014, Wang2018}. Conservative estimates of these elemental abundances and the mineralogy of rocky planets at varied orbital distances are presented in Section \ref{sec:conserve}. It should be noted that these points are not observational \textquoteleft data' but indications of the model-based devolatilisation scales (i.e. planet-to-host abundance ratios) of these elements versus condensation temperatures ($T_C$). These devolatilisation scales are also numbered in the second and third columns of Table \ref{tab:abu}, in \% (linear) and in dex (logarithmic), respectively.
The normalisation-reference element is Al, the most refractory major element. %
Normalising to other refractory elements (e.g. Ca) will not affect the analysis or results, since this normalisation will not change the relative abundance ratio of other elements to oxygen that is the key to determining the redox state of a planet.
It is worth noting that the $T_C$ in Figure \ref{fig:model} refers to 50\% condensation temperatures of \cite{Lodders2003}, except for C and O. As discussed in \cite{Wang2018b}, an effective condensation temperature of C and O under the assumption of \textit{non-equilibrium multiphase condensation} is a better indication of the volatility of the two elements, than 50\% condensation temperature of them under the assumption of \textit{equilibrium single-phase condensation} in \cite{Lodders2003}. We use their best estimates of the effective $T_C$ for C and O: 305 K and 875 K respectively.

\subsection{Chemical network of the mantle of a terrestrial planet}
\label{sec:mantle}
In this study, the mantle of a terrestrial planet is limited to silicate mineralogy. This limitation is appropriate for two reasons: i) our devolatilisation algorithm that is established for the Sun to a silicate-mantle terrestrial planet is unlikely to be applicable for a planetary system with C/O ratio larger than 0.8 \citep[and thus forming carbide planets -- see ][]{Bond2010b}; ii) More recent studies \citep[e.g.][]{Fortney2012, Nissen2013, Teske2014, Brewer2016} have suggested that prior studies \citep[e.g.][]{Bond2010b, Delgado2010, Petigura2011} overestimated C/O ratios, namely carbide planets may not be as abundant as previously thought. 

We assume that the composition of silicate mantle varies within the chemical system of SiO$_2$-CaO-Na$_2$O-MgO-Al$_2$O$_3$-FeO-NiO-SO$_3$. And it is similar to the NCFMAS (Na$_2$O-CaO-FeO-MgO-Al$_2$O$_3$-SiO$_2$) mantle model adopted in \citet{Dorn2015}, but the oxides are reordered in the oxidation sequence (or ease of oxidation) \citep{Johnson1999}.
NiO and SO$_3$ are also added into the system. Due to the strong affinity of Ni with Fe, NiO is always expected if FeO is present in the mantle. 
SO$_3$ can combine with other oxides to form sulfate compounds (e.g. CaSO$_4$), which are important minerals in the silicate mantle (though not as abundant as silicate compounds). The additions of NiO and SO$_3$ will also make estimates of the partition of Ni and S into the core more accurate (see Section \mbox{\ref{sec:core})}. Sulfide minerals (except for FeS) are the major host of chalcophile (sulfur-loving) elements, but they are not taken into account in our mantle system since the abundances of all chalcophile elements (e.g. Cu, Zn, and Ag) are negligible in comparison with the abundances of the major-rock forming elements considered here.

With regard to a planet's mantle temperature ($T$), pressure ($P$), and oxygen fugacity ($f$O$_2$, quantifying the oxidation potential of a system), the mineral host of carbon can be either oxidised carbonate species (denoted as C$^{4+}$ or CO$_2$ in an oxidised form) \citep{Panero2008, Boulard2011} or the reduced native element (e.g. graphite/diamond) \citep{Walter2008, Dasgupta2010}. Oxidised carbon will be considered only when all major elements listed in the chemical network of the mantle have been oxidised, since carbon is experimentally shown to be oxidised later than Fe (presumably Ni as well) over the entire pressure and temperature ranges of Earth's mantle \citep{Unterborn2014}. We also assume that if oxygen fugacity is extremely limited, metals like Ca, Na, Mg and Al can be present natively in the reduced environment of the mantle, while Fe and Ni will be all partitioned into the core. 

\subsection{Chemical network of the core of a terrestrial planet}
\label{sec:core}
\begin{figure}
	\includegraphics[trim=0.5cm 2.0cm 1.0cm 1.5cm, scale=0.32, angle=90]{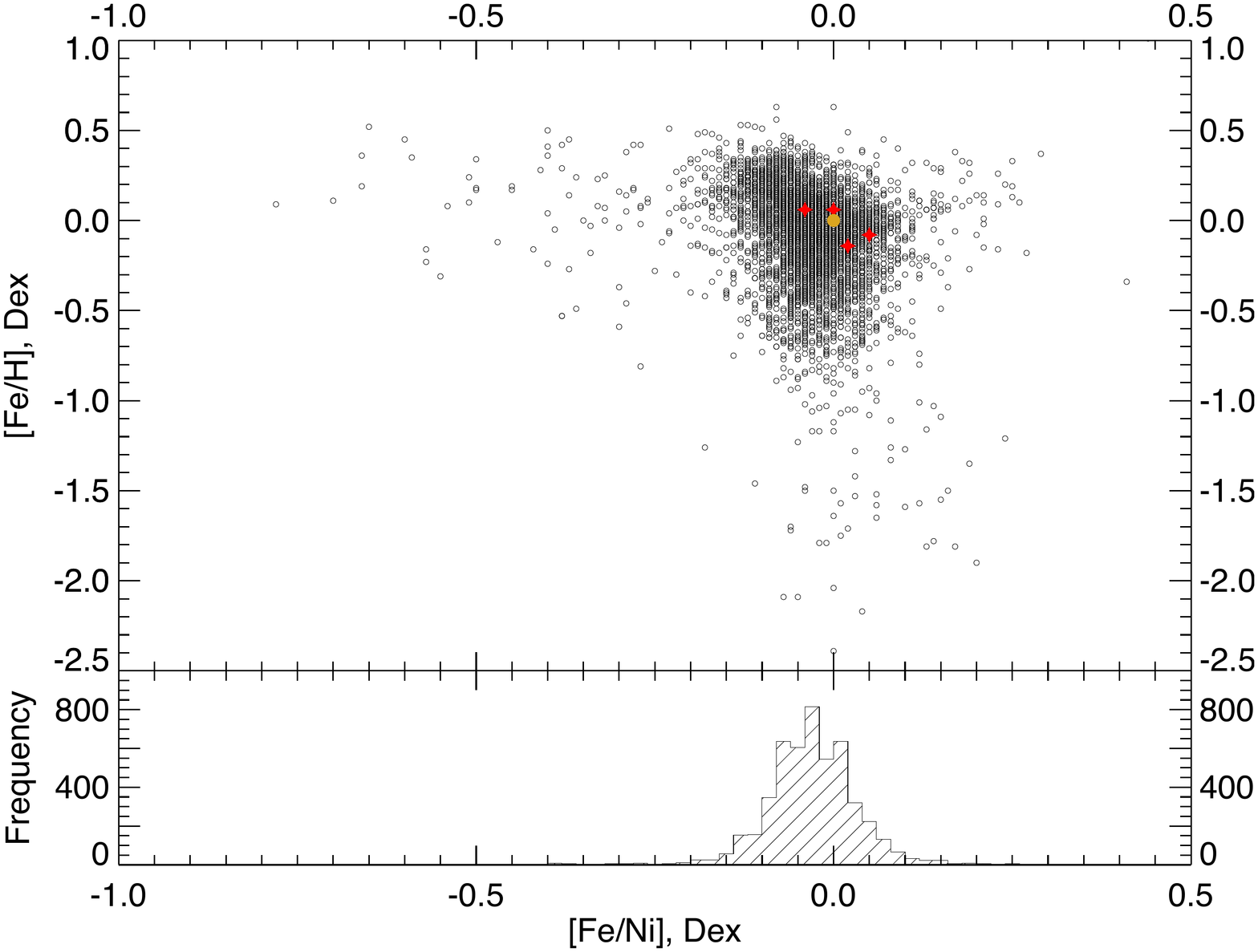}
	\caption{Distribution of [Fe/Ni] of more than 4900 FGK-type stars within 150 pc of the Sun. The values of [Fe/Ni] and [Fe/H] are relative to the Sun \citep{Asplund2009}, drawn from the Hypatia Catalog \citep{Hinkel2014}. A Gaussian fit to the distribution of [Fe/Ni] gives [Fe/Ni] = -0.033 $\pm$ 0.049 dex (relative to [Fe/Ni]$_{\sun}$ = 1.28 dex). 
	The corresponding value of Fe/Ni in linear is 17.7 ($=10^{-0.033+1.28}$) with 1$\sigma$ uncertainty of $\sim$ 2.0.
	A conservative value of $18\pm4$ is adopted to constrain the Fe/Ni ratio in the core of a terrestrial exoplanet (see the main text). The yellow point represents the Sun while the red filled pluses indicate our case stars.}
	\label{fig:Fe2Ni}
\end{figure}

\begin{figure*}
	\includegraphics[trim=1.0cm 2.5cm 0.5cm 2.0cm, scale=0.6,angle=90]{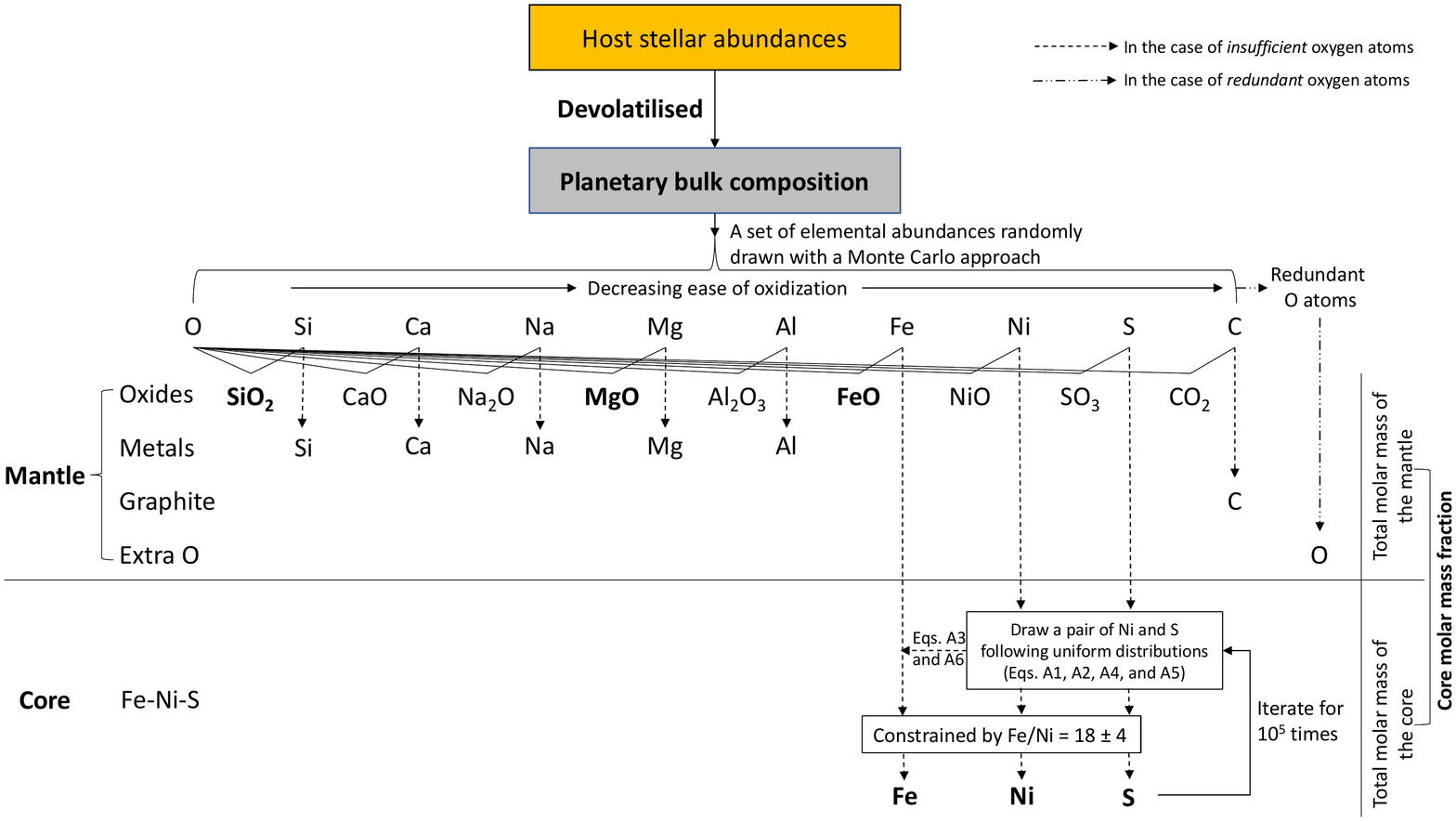}
	\caption{Computational procedure scheme of elemental fractionation between the mantle and the core of a terrestrial planet. The detailed descriptions of the procedure can be found in Appendix \ref{app:fractionation}.}
	\label{fig:procedure}
\end{figure*}
We assume that a terrestrial planet's core consists of the Fe-Ni-S alloy. Nickel is added into the core's constituents as it has a similar siderophile tendency as iron. Sulfur is a leading candidate for the principal light element in the core because it has a strong affinity for iron \citep{Li2014}, and indeed evidence for large-scale sulfide fractionation during Earth's mantle-core differentiation has been found \citep{Savage2015}. This does not preclude the existence of other light element candidates like silicon and oxygen in the core, but their fractionation between the mantle and the core is very uncertain without knowing the actual internal pressure and temperature of such a planet \citep{Hirose2017}. Therefore, in this study, we limit the light element in a terrestrial planet's core to be sulfur only. Furthermore, sulfur is capable of reducing the core's melting temperature, density, and surface tension \citep{Li2014}, which may be important to the dynamo modelling of planetary magnetic fields in further studies. 

It is important then to put constraints on the fractionation of these elements between the mantle and core. The cosmic ratio of Fe/Ni is $17.4\pm0.5$ (by mass) 
as drawn from a variety of chondrites in the solar system \citep{McDonough2017}. This value could vary from star to star, but the variance should be small based on the nucleosynthesis of the two elements in stars. Indeed, by analysing the Fe/Ni ratios of more than 4900 FGK-type stars within 150 pc of the Sun from Hypatia Catalog \citep{Hinkel2014}, we have found that the number ratio of Fe/Ni is 17.7 with 1$\sigma$ uncertainty of $\sim$ 2.0 (see Figure \ref{fig:Fe2Ni}).
Although their ratio
in the planetary core might not be fixed, it will deviate little from the cosmic ratio considering their very similar refractory and siderophile features (William F. McDonough, personal communication). 
Indeed, a recent metal-silicate partitioning experiment has shown the ideality of the Fe/Ni ratio at conditions similar to Earth's core \citep{Huang2018}.
To be conservative, a value of 18 $\pm$ 4 (by number) is adopted to constrain the Fe/Ni ratio in the core of a terrestrial planet. This adoption is appropriate for this study as the upper limit (22) of this value has covered the maximum value of Fe/Ni of our selected planet hosts. The lower limit (14) is $\sim1.5\sigma$ lower than the Fe/Ni ratios of our selected cases, but it is a good assumption as Fe/Ni in the core of a terrestrial planet may go lower (not higher) than its value in the bulk planet
\citep{Ringwood1986, Seifert1988, McDonough1995, Wang2018}.
%(Hugh O'Neill, personal communication).

Another constraint is the abundance range of Ni and S that may be present in the core. The upper limits of Ni and S in the core can be up to their abundances in the bulk planet, referring to the scenario of the Earth: $>90\%$ of Ni and $>95\%$ of S are in the Earth's core \citep{McDonough2014, Wang2018b}. The lower limits can be assumed as nil, such as in the extreme scenario that the oxygen fugacity is too high and all metals including Fe could be fully oxidised (then a planet with no core would result) \citep{Elkins2008}. 
See more details in Section \ref{sec:analysis} and Appendix \ref{app:fractionation}.

\subsection{Analysis}
\label{sec:analysis} 
Our sample of planet host stars includes Kepler-10 (K10), Kepler-20 (K20), Kepler-21 (K21), and Kepler-100 (K100), each of which has been confirmed to host at least one super-Earth (with 3-10 Earth masses), with high-precision host stellar abundances (available for almost all ten elements indicated in Figure \ref{fig:model} while for at least Mg, Si, Fe and O).
For K10, we have two sets of stellar elemental abundances: \cite{Santos2015} and \cite{Liu2016}. For other selected cases, our stellar elemental abundances are from \cite{Schuler2015} (with a typical uncertainty of $\lesssim$ 0.04 dex). The selection of these case stars in this work does not exclude the applicability of our method to other planet hosts in a more extensive case study in the future.

Following the recommended constraint on planetary bulk composition in Sect. \ref{sec:bulk}, we apply the devolatilisation pattern of \cite{Wang2018b} to the elemental abundances of the planet hosts of our sample. This leads to the first-order estimates of bulk elemental composition of \textit{potential} rocky exoplanets that could be within the habitable zone around these host stars (see Table \ref{tab:abu}). 
For brevity and convenience, we call such an exoplanet \textquoteleft exo-Earth' (or \textquoteleft exoE' when it is preceded by the name of its host star). However, this term means nothing about the similarity of such a planet to the Earth, except for the connotation that this planet is derived from its host star based on the equivalent devolatilisation scale from the solar nebula to the Earth (the limitations of this assumption are discussed in Section \ref{sec:limits}). The uncertainties on the bulk compositions of these exo-Earths in Table \ref{tab:abu} are propagated from the 1$\sigma$ uncertainty of the devolatilisation pattern of \cite{Wang2018b} and the uncertainties associated with the elemental abundances of the corresponding planet hosts.

We analyse the key elemental ratios (e.g. Mg/Si, Fe/Si, and C/O) that modulate the primary mineralogy of a terrestrial planet. Mg/Si influences the rock types of a silicate mantle. A higher Mg/Si indicates a more olivine-dominated mantle, otherwise a more pyroxene-dominated one. Pyroxene can accommodate more water than olivine, and pyroxenites have higher capacity to hold water.
Fe/Si may approximate the zeroth-order core mass fraction of a planet, upon the oxidation state of the planet.
C/O is critical to determine the oxygen fugacity and thus the oxidation state of a planet.
The errors in these key elemental ratios are calculated using a Monte Carlo approach by drawing $2\times10^4$ values of the estimated planetary bulk composition for each element following a Gaussian distribution around the uncertainties. 

Based on the recommended chemical networks for the mantle and core of a terrestrial planet in Sects. \ref{sec:mantle} and \ref{sec:core}, the elemental fractionation between the mantle and the core is performed using the stoichiometric balance between the budget of oxygen atoms in the bulk planet and the abundances of oxides/compounds to be considered for the planet. The computational procedure is summarised in Figure \ref{fig:procedure}, while a detailed description can be found in Appendix \ref{app:fractionation}.

As a verification for this set of computations, we apply this procedure to the proto-Sun \citep{Wang2018b}. 
The predicted planetary interior compositions of the model Earth in Table \ref{tab:verify} are consistent (within uncertainties) with the independent estimates of the composition of the pyrolite silicate Earth \citep{McDonough1995}, the composition of Earth's core \citep{McDonough2014}, as well as the seismologically-constrained core mass fraction \citep{Wang2018}.

\begin{table}
	\caption{Comparison of the estimates of the interior compositions of the model Earth (as devolatilised from the proto-Sun$^a$) with other independent estimates}
	\begin{center}
		\begin{tabular}{ll  l c}
			\toprule
			& \multicolumn{1}{l}{Quantity} & Model Earth & \cite{McDonough1995}\\
			\hline 
			\multirow{12}{*}{\rotatebox[origin=c]{90}{Mantle}} & SiO$_2$ &50.3 $\pm$ 4.7 &45.0\\	
			& CaO& 3.60 $\pm$ 0.34 &3.55\\	
			& Na$_2$O& $0.44\pm0.04$ &0.36\\
			& MgO& 38.0 $\pm$ 6.2 &37.8\\
			& Al$_2$O$_3$& 4.40 $\pm$ 0.40 &4.45\\
			\vspace{1mm}
			& FeO &$1.87^{+7.14}_{-1.87}$ &8.05\\
			\vspace{1mm}
			& NiO&$0.40^{+0.45}_{-0.40}$ &0.25\\
			& SO$_3$&0.67 $\pm$ 0.62 &\textendash\\
			& CO$_2$&\textendash &\textendash\\
			& \multicolumn{1}{p{1.0cm}}{Graphite}&$0.39\pm0.07$ &\textendash\\
			& Metals&\textendash &\textendash\\
			& Extra O&\textendash &\textendash\\
			\hline
			&& &\cite{McDonough2014}$^b$\\
			\multirow{4}{*}{\rotatebox[origin=c]{90}{Core}} & Fe &$93.28\pm1.23$ & 92.33 \\
			& Ni& 5.22 $\pm$ 0.63 &5.62 \\ 
			& S& 1.50 $\pm$ 1.05 &2.05 \\
			\hline
			
			\multicolumn{2}{p{2cm}}{Core mass fraction (wt\% planet)} &$31.3\pm5.3$ &$32.5\pm0.3$ $^c$\\
	
			\bottomrule
			%%%add the footnotes				
			\multicolumn{4}{p{8cm}}{$^a$ The protosolar abundances are from \cite{Wang2018b}.} \\
			\multicolumn{4}{p{8cm}}{$^b$ Mass fractions of the three elements in the core of \cite{McDonough2014} have been renormalised under the assumption of only the three elements present in the core as practiced in this study.} \\
		    \multicolumn{4}{p{8cm}}{$^c$ Refers to \cite{Wang2018}, in which the core mass fraction is integrated from Earth's radial density profiles.} \\				
		\end{tabular}
	\end{center}
	\label{tab:verify}
	%	}
\end{table}

\section{Results}
\label{sec:results}
\subsection{Estimates of key elemental ratios}
\label{sec:ratios}
\begin{figure*}
	\includegraphics[trim=5.0cm 2.0cm 3.0cm 1.0cm, scale=0.65,angle=90]{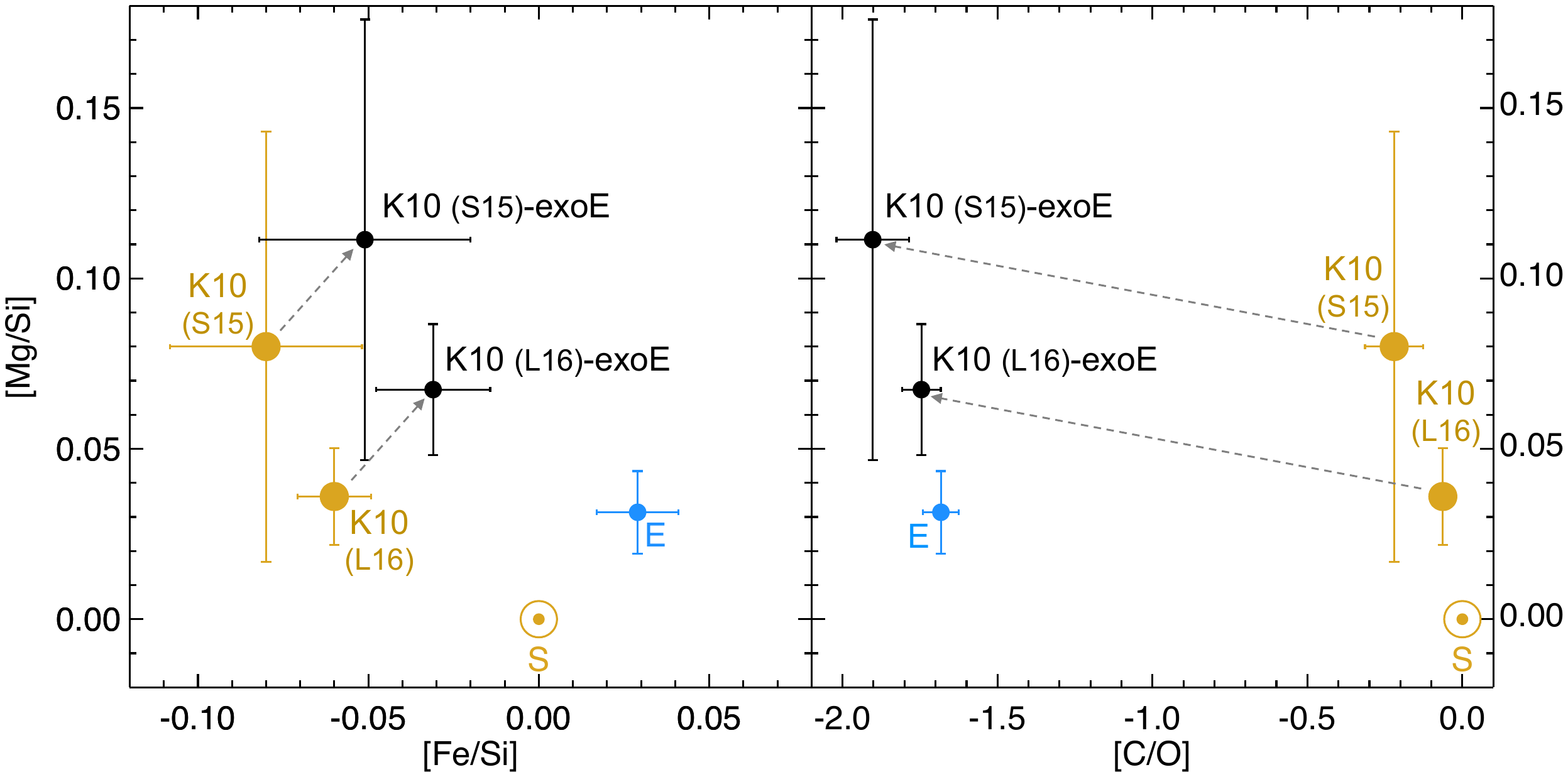}
	\caption{In the case of Kepler-10, the comparison of key elemental ratios ([Mg/Si], [Fe/Si], and [C/O], in dex) between the host star and its potential exo-Earth. This comparison includes two sets of host stellar abundances: \protect\cite{Santos2015} (labeled as \textquotedblleft K10 (S15)") and \protect\cite{Liu2016} (labeled as \textquotedblleft K10 (L16)"), and their corresponding exo-Earths are potted as smaller black dots, labeled as \textquotedblleft K10 (S15)-exoE" and \textquotedblleft K10 (L16)-exoE", respectively. 
	Key elemental ratios in the Sun (\textquoteleft S', the symbol $\odot$ in yellow, \protect\cite{Asplund2009}) and in Earth that is devolatilised from it (\textquoteleft E', the dot in blue) are plotted as references. The dashed arrows indicate the trajectories of these elemental ratios from Kepler-10 to K10-exoE due to the applied devolatilisation.}
	\label{fig:k10b}
\end{figure*}
\begin{figure*}
	\includegraphics[trim=3.0cm 1.5cm 4.0cm 0.8cm, scale=0.6,angle=90]{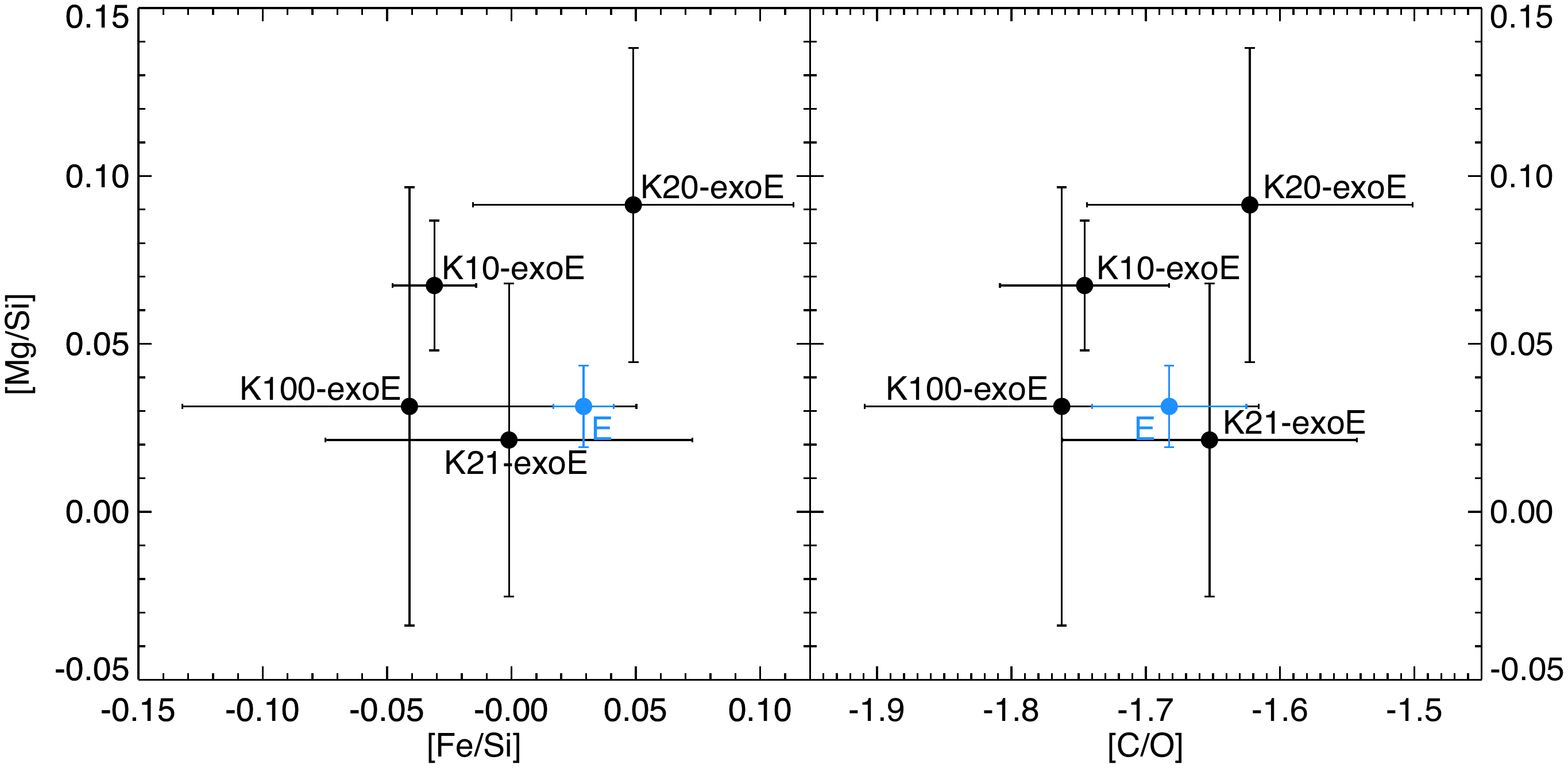}  
	\caption{For all studied cases of planet hosts, the comparison of these key elemental ratios between their potential \textquotedblleft exo-Earths", labeled as K10-exoE, K20-exoE, K21-exoE, and K100-exoE, which are respectively corresponding to the potential habitable-zone terrestrial exoplanets orbiting the planet hosts: Kepler-10, Kepler-20, Kepler-21, and Kepler-100. Host stellar abundances are all from \protect\cite{Schuler2015}, except for Kepler-10 that is from \protect\cite{Liu2016}. The Earth (same as Figure \ref{fig:k10b}) is plotted as a reference.}
	\label{fig:ratios}
\end{figure*}

In Figure \ref{fig:k10b}, we compare the key elemental ratios between the host Kepler-10 and its potential exo-Earth, by using two sets of host stellar abundances: one is with a typical uncertainty of $\gtrsim$ 0.06 dex \citep{Santos2015} and the other is with a typical uncertainty of $\lesssim$ 0.02 dex \citep{Liu2016}. The planetary elemental ratios are derived from the \textquotedblleft devolatilised\textquotedblright\thickspace host stellar abundances -- i.e. the predicted planetary bulk composition listed in Table 1. With the more precise host stellar abundances, the planetary elemental ratios for Mg/Si, Fe/Si, and C/O are all significantly different from the corresponding host stellar elemental ratios. Namely, they do not overlap within their uncertainties. Whereas, these significant differences are substantially dismissed with the less precise host stellar abundances, except for C/O (it difference is up to 1.7 dex, i.e., a factor of $\sim$ 50). This is the first attempt (based on theoretical models) to compare the key elemental ratio differences between a terrestrial exoplanet and its host star, rather than those differences of an extrasolar star from the Sun as usually done in the literature \citep[e.g.][]{Bond2010b, Santos2015, Brewer2016}. Nonetheless, in order to distinguish two unique terrestrial exoplanets (orbiting different stars), high-precision host stellar abundances are required, as concluded in \cite{Hinkel2018} as well. In our subsequent modelling of planetary interiors, the less-precise Kepler-10 abundance estimates (also available for fewer elements) of \cite{Santos2015} are therefore no longer used.

We also compute the key elemental ratios for potential exo-Earths orbiting the other three planet hosts (e.g. Kepler-20, Kepler-21, and Kepler-100), based on the \textquotedblleft devolatilised\textquotedblright\thinspace host stellar abundances listed in Table \ref{tab:abu}. The key elemental ratios of these exo-Earths are plotted in Figure \ref{fig:ratios}. 
When considering the deviations of these exoplanetary key elemental ratios (including the uncertainties) from Earth's, K21-exoE is overall the most Earth-like
while K10-exoE is the least.
If only based on Mg/Si (i.e. modulating the rock types of a silicate mantle), 
K100-exoE and K21-exoE are more Earth-like than K10-exoE and K20-exoE. 
These similarities will be refined in further by the following modelling of planetary interiors.

\subsection{Estimates of planetary interiors}
\label{sec:interior}
%
%% Example table
\begin{table*}
	\caption{Estimates of the mantle and core composition as well as core mass fraction of potential habitable-zone terrestrial exoplanets (i.e. exo-Earths) orbiting the studied host stars: Kepler-10 (K10), Kepler-20 (K20), Kepler-21 (K21), Kepler-100 (K100).}
	\begin{center}
		\begin{tabular}{ll  llll lc}
			% {\tiny
			%\hline
			\toprule
			& \multicolumn{1}{l}{Quantity} &
			\multicolumn{4}{c}{Potential habitable-zone terrestrial exoplanets (i.e. exo-Earths)$^a$} && Pyrolite Silicate Earth\\
			
			\cline{3-6}
			& (molar wt\%)  & K10-exoE & K20-exoE & K21-exoE & K100-exoE &&\cite{McDonough1995}\\
			\hline 
			\multirow{12}{*}{\rotatebox[origin=c]{90}{Mantle}} & SiO$_2$ &$30.3\pm2.6$&$50.7\pm3.3$&$49.9\pm4.9$&$40.5\pm4.4$ &&45.0\\	
			& CaO &$2.50\pm0.22$&$4.39\pm0.34$&$3.85\pm0.38$&$2.98\pm0.36$ &&3.55\\	
			& Na$_2$O &$0.23\pm0.02$&$0.43\pm0.03$&$0.41\pm0.04$&$0.41\pm0.05$ &&0.36\\
			& MgO&$29.2\pm2.4$&$25.7\pm3.6$&$43.3\pm3.9$&$36.0\pm4.0$ &&37.8\\ 
			& Al$_2$O$_3$&$3.28\pm0.31$&\textendash&\textendash&$4.01\pm0.47$ &&4.45\\
			\vspace{1mm}
			& FeO&$31.3\pm5.1$&\textendash&$1.53^{+8.35}_{1.53}$&$13.9\pm7.9$&&8.05\\ 
			\vspace{1mm}
			& NiO&$1.71\pm0.34$&\textendash&$0.15^{+0.40}_{-0.15}$&$0.95\pm0.51$ &&0.25\\ 
			& SO$_3$&$1.25\pm0.46$&\textendash&$0.50\pm0.48$&$0.90\pm0.51$ &&\textendash\\ 
			& CO$_2$&\textendash&\textendash&\textendash&\textendash &&\textendash \\
			& \multicolumn{1}{p{1.0cm}}{Graphite}&$0.28\pm0.05$&$0.38\pm0.05$&$0.42\pm0.06$&$0.30\pm0.07$ &&\textendash\\ 
			& Metals&\textendash&$18.4\pm10.7$&\textendash&\textendash &&\textendash\\
			\hline
			& &&&& &&\cite{McDonough2014}$^b$\\ 
			\multirow{4}{*}{\rotatebox[origin=c]{90}{Core}} & Fe &$85.7\pm5.0$&$94.5\pm0.6$&$94.1\pm1.3$&$92.9\pm1.9$ && 92.33 \\
			& Ni&$5.25\pm0.73$&$5.49\pm0.63$&$4.76\pm0.65$&$5.69\pm0.68$ &&5.62\\  
			& S&$9.07\pm5.24$&\textendash&$1.16\pm1.14$&$1.40^{+1.83}_{-1.40}$ &&2.05\\
			\hline
			
			\multicolumn{2}{p{2cm}}{Core mass fraction (wt\% planet)} &$1.4^{+5.0}_{-1.4}$&$35.3\pm5.1$&$31.9\pm5.9$&$19.6\pm6.3$ &&$32.5\pm0.3$ $^c$\\

			\bottomrule
			%%%add the footnotes	
			\multicolumn{8}{p{13cm}}{$^a$ Bulk elemental compositions of these exo-Earths are from Table \ref{tab:abu}.} \\			
			\multicolumn{8}{p{13cm}}{$^b$  Mass fractions of the three elements in the core of \cite{McDonough2014} have been renormalised under the assumption of only the three elements in the core.}\\		
			\multicolumn{8}{p{13cm}}{$^c$  Refers to \cite{Wang2018}, in which the core mass fraction is integrated from Earth's radial density profiles.}			
		\end{tabular}
	\end{center}
	\label{tab:comp}
	%	}
\end{table*}

\begin{figure*}
	\includegraphics[trim=2.5cm 2.5cm 2.2cm 2.5cm, scale=0.73,angle=90]{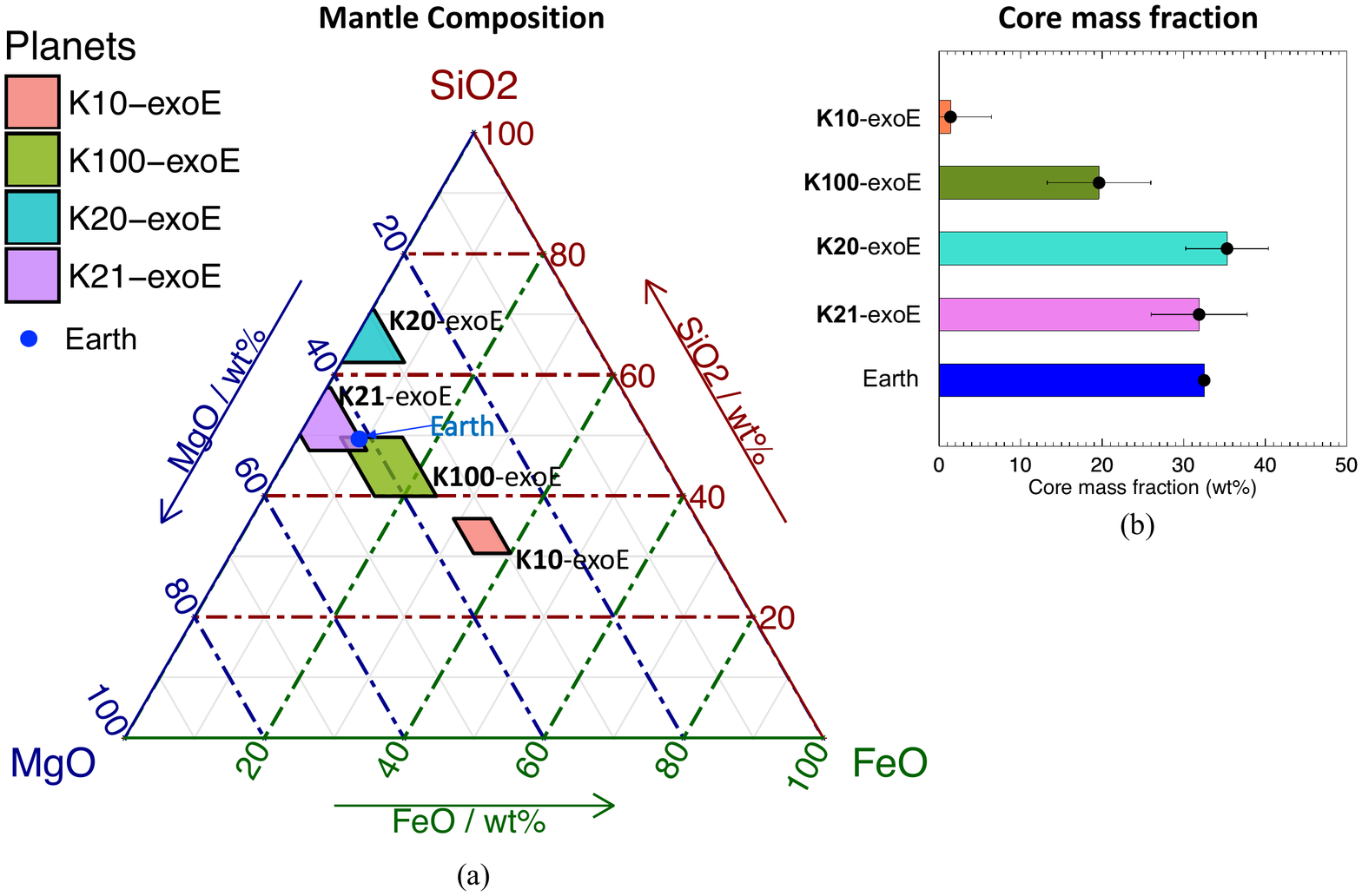}
	\caption{(a) A ternary diagram illustrating the estimates of the mantle composition (normalized by SiO$_2$ + MgO + FeO = 100 wt\%, adapted from Table \ref{tab:comp}.) for potential exoEarths orbiting the studied host stars. The Earth's mantle composition of SiO2, MgO, and FeO in \protect\cite{McDonough1995} is normalised similarly and plotted as a reference. (b) A bar graph comparing the estimates of the core molar mass fraction of these exoEarths (see Table \ref{tab:comp}.). The Earth's core mass fraction (32.5 $\pm$ 0.3 wt\%) in \protect\cite{Wang2018} is plotted as a reference. For Kepler-10, only one of the two sets of results for its potential exoEarth is plotted here, which refers to the higher-precision estimates of the host stellar abundances \protect\citep[i.e.][]{Liu2016}.}
	\label{fig:results}
\end{figure*}

Based on the recommended chemical networks of the mantle and core of a terrestrial planet, we estimate the mantle and core composition as well as the core mass fraction for these potential exo-Earths. These estimates are listed in Table \ref{tab:comp}. We also extract the information of the major components (SiO$_2$, MgO, and FeO) in the mantle and of the core mass fraction, and then compare them in Figure \ref{fig:results}. It should be noted that the mantle composition in the ternary diagram of Figure \ref{fig:results} has been renormalized by ignoring all minor components and assuming SiO$_2$ + MgO + FeO = 100 wt\%. 

From Table \ref{tab:comp} and Figure \ref{fig:results}, we can further identify that K21-exoE is the most Earth-like, in terms of all interior parameters we have modeled, including the mantle composition, core composition and core mass fraction. This similarity has been implied by the analyses of key elemental ratios above, but it is more conclusive and obvious here. In particular, according to the \textit{normalized} mantle composition in the ternary diagram of Figure \mbox{\ref{fig:results}}, K20-exoE has the highest SiO$_2$ ($66.4\pm4.3$ wt\%) and the lowest MgO ($33.6\pm4.7$ wt\%). The mantle rocks of K20-exoE therefore will be more enriched in pyroxene (MgSiO$_3$) compared to other planet cases studied here (including the Earth). In contrast, K10-exoE has the lowest SiO$_2$ ($33.4\pm2.9$ wt\%) and the equivalent MgO ($32.2\pm2.8$ wt\%), and thus its mantle rocks will potentially be more forsterite (Mg$_2$SiO$_4$) enriched than the others. For K21-exoE and K100-exoE, there is an overlap of their mantle compositions with that of the Earth as shown in the ternary diagram, so they would have similar mantle rocks as the Earth's. However, as shown in the bar diagram of core mass fractions (Figure \mbox{\ref{fig:results}}b), K100-exoE has a relatively smaller core (in terms of core mass fraction) than K21-exoE and the Earth, and the latter two are comparable. This can be explained by the differences of planetary bulk compositions (Table \mbox{\ref{tab:abu}}), in which the mean of Fe abundance of K100-exoE ($-0.17\pm0.08$ dex) is $\sim 1\sigma$ lower than that of K21-exoE ($-0.09\pm0.07$ dex) and that of the Earth ($-0.07\pm0.02$ dex) while their abundances of other major elements including O, Mg, and Si are similar. %within their uncertainties.} 

In addition, FeO is the most enriched in the mantle of K10-exoE (while it is the least (nil) in the mantle of K20-exoE). Namely, in comparison with other planet cases studied here, the differentiation of iron into the core of K10-exoE is the least efficient, thus leading to the smallest core (0--6.4 wt\%) among the studied cases. The reason is present in its distinct host stellar abundances. Comparing with other selected planet hosts, Fe abundance in Kepler 10 \citep[refering to][]{Liu2016} is about 1--3$\sigma$ lower while its O abundance relative to Mg and Si (i.e. O - Mg - 2Si, in dex) is 1--5$\sigma$ higher.  
As a consequence, almost all Fe in Kepler 10 is oxidised, and only a tiny amount of metallic Fe sinks into the core, thus leading to the substantially smaller core mass fraction in K10-exoE as modelled. This reveals the importance of the trade-off between oxygen and other major rock-forming elements to the exoplanetary interior modelling.

\subsection{Conservative estimates}
\label{sec:conserve}
\begin{table*}
	%	{\footnotesize
	%\centering
	\caption{Conservative estimates of the mantle and core composition as well as core mass fraction of any rocky exoplanet orbiting the studied host stars}
\begin{center}
	\begin{tabular}{ll  llll p{0.5cm} llll}
		% {\tiny
		%\hline
		\toprule
		& \multicolumn{1}{l}{Quantity} & %\multicolumn{1}{c}{} & 
		\multicolumn{4}{c}{Less depleted$^a$} & &\multicolumn{4}{c}{More depleted$^b$}\\
		%& \multicolumn{6}{c}{Starting with the host stellar elemental abundances} \\
		\cline{3-6} \cline{8-11}
		& (molar wt\%)  & \multicolumn{1}{l}{K10} & K20 & K21 & K100 & & \multicolumn{1}{l}{K10}& K20 & K21 & K100 \\
		\hline
		
		\multirow{12}{*}{\rotatebox[origin=c]{90}{Mantle}} & SiO$_2$ &$23.3$ &$26.4$&$25.3$&$24.0$& 
		&$51.0$ &$47.5$&$61.0$&$59.1$\\	
		& CaO &$1.68$ &$2.06$&$1.76$&$1.56$& 
		&$4.96$ &\textendash&\textendash&$5.09$\\	
		& Na$_2$O &$0.25$ &$0.32$&$0.31$&$0.34$&
		&$0.27$ &\textendash&\textendash&0.41\\
		& MgO&$22.4$ &$27.6$&$22.5$&$21.4$&
		&$26.5$ &\textendash&\textendash&0.55\\
		& Al$_2$O$_3$ &$2.17$ &2.31&\textendash&$2.00$& 
		&\textendash &\textendash&\textendash&\textendash\\
		& FeO&$25.1$&37.2&$32.5$&$27.5$&
		&\textendash &\textendash&\textendash&\textendash\\
		& NiO&$1.33$ &1.90&$1.48$&$1.51$&
		&\textendash &\textendash&\textendash&\textendash\\
		& SO$_3$&$2.69$ &\textendash&$2.25$&$2.40$&
		&\textendash &\textendash&\textendash&\textendash\\
		& CO$_2$&4.90 &1.07&5.02&4.35&
		&\textendash&\textendash&\textendash&\textendash\\
		& \multicolumn{1}{p{1.0cm}}{Graphite}&\textendash &1.01&\textendash&\textendash&
		&0.08 &0.07&0.08&0.07\\
		& Metals&\textendash &\textendash&\textendash&\textendash&
		&17.2 &$52.4$&$38.9$&$34.8$\\
		& Extra O&16.2 &\textendash&8.90&14.9&
		&\textendash &\textendash&\textendash&\textendash\\
		\hline

		\multirow{4}{*}{\rotatebox[origin=c]{90}{Core}} & Fe &\textendash &\textendash&\textendash&\textendash& 
		&93.8 &94.0&93.9&92.5\\
		& Ni&\textendash &\textendash&\textendash&\textendash&
		&5.28 &6.05&5.39&6.68\\
		& S&\textendash &\textendash&\textendash&\textendash&
		&0.92 &\textendash&0.69&0.86\\
	
		\hline
		
		\multicolumn{2}{p{2cm}}{Core mass fraction (wt\% planet)} &0 &0&0&0&
		&$31.7$ &$38.3$&$36.1$&$32.8$\\

		\bottomrule
		%%%add the footnotes				
		\multicolumn{11}{p{11cm}}{$^a$ Applying the upper bound of 3$\sigma$ uncertainty of the Sun-to-Earth devolatilisation pattern to the upper limits of host stellar abundances.} \\		
		\multicolumn{11}{p{11cm}}{$^b$ Applying the lower bound of 3$\sigma$ uncertainty of the Sun-to-Earth devolatilisation pattern to the lower limits of host stellar abundances.}
						
	\end{tabular}
\end{center}
\label{tab:comp-lmt}
%	}
\end{table*}

We also apply the upper and lower limits of the 3$\sigma$ error bar of the best-fitting devolatilisation pattern of \cite{Wang2018b} to the respective upper and lower limits of the host stellar elemental abundances. It should be noted that the 3$\sigma$ is conveniently amplified over the uncertainty associated with the coefficients of the pattern; thus, the resultant 3$\sigma$ range is more conservative than a range constrained by a 3$\sigma$ chi-square fit. The resultant conservative estimates -- \textquotedblleft less depleted\textquotedblright and \textquotedblleft more depleted\textquotedblright -- are listed in Table \ref{tab:comp-lmt}.
 
Though the volatile gradient of a planetary system versus the distance to its central star is still controversial \citep{Morgan1980, Palme2000, Wang2016, Jin2018}, the dichotomy of planets from \textquoteleft rocky' to \textquoteleft gas/icy' or from  \textquoteleft warm' to \textquoteleft cold' would be expected from the inward to outward in a planetary system (if those planets exist around the central star).
Therefore, the \textquotedblleft less depleted\textquotedblright case might assemble the interior properties of \textquoteleft cold' rocky bodies (e.g. asteroids) beyond the outer edge of the habitable zone but within the snowline of a planetary system. In contrast, the \textquotedblleft more depleted\textquotedblright case might be a better proxy %than the estimates in Table \ref{tab:comp}, 
for the interior properties of \textquoteleft warm' rocky planets, such as the known planets Kepler-10b, Kepler-20b, Kepler-21b, and Kepler-100b, which are closer to their host stars. 

\section{Discussion}
\label{sec:disc}
\subsection{Comparison with previous studies}
The diversity of planetary interiors should be expected. Neither our conservative estimates in Table \ref{tab:comp-lmt} nor the estimates for habitable zone terrestrial exoplanets in Table \ref{tab:comp} should be taken as an explicit interpretation of the interiors of any existing planet orbiting these studied host stars, but only the plausible ranges constrained by different scenarios as demonstrated in the paper.
The previous studies \citep[e.g.][]{Santos2015, Weiss2016, Brugger2017} usually state that their analyses are for an explicit planet around a parent star but this statement should be taken with caution. For example, the core mass fraction (wt\%) of Kepler-10b has been previously estimated to be $27.5\pm1.7$ \citep{Santos2015}, $17\pm11$ \citep{Weiss2016}, and 10-33 \citep{Brugger2017}, which seem to approximate our \textquotedblleft more depleted\textquotedblright case (30) for a \textquoteleft warm' rocky planet orbiting Kepler 10. Whereas, those previous estimates are achieved by fixing all Fe into the core \citep{Santos2015} or relying on Mg\# (= Mg/(Mg + Fe)) \citep{Weiss2016, Brugger2017}, without considering oxygen fugacity/budget that controls the oxidation state of a planet.
If one generalises those fixed assumptions to potential rocky planets at different distances to the same star, the same chemistry and interior structure for these planets would result. Although taking into account planetary mass and radius, the modelling results could be more specific to such planets, but the results would still be degenerate to the degree to which oxygen fugacity/budget would vary in the disc. When targeting a potentially habitable planet for further characterisation with future missions, we should particularly be cautious with this kind of \textquoteleft explicit' claims, which may only assemble one scenario of possible interior properties of rocky exoplanets orbiting at different distances to their host stars. The depletion of oxygen in Kepler-10 planets relative to the host star is not considered in \cite{Dorn2017}, but they assume Fe/Si$_{\textrm{mantle}}$ = [0, Fe/Si$_{\textrm{star}}$] (uniform distribution), implicitly resulting in varied differentiation of Fe into the core, which somehow assembles the consideration of oxygen fugacity/budget in determining the fractionation of Fe between the mantle and core.

Complex mineral compounds (e.g. (Mg,Fe)SiO$_3$, (Mg,Fe)$_2$SiO$_4$) have been presumed to be present in the mantle rocks of super-Earths in the studies of \citet{Santos2015}, \cite{Weiss2016}, \cite{Brugger2017}, and \cite{Hinkel2018}. However, for exoplanets that are larger than Earth, their mineralogies may not be sensibly interpretable in using the Equation-of-State (EoS) parameters measured with respect to the pressure and temperature regime of the Earth \citep{Mazevet2015}. Therefore, we have chosen to use first-order mineral forms (i.e. oxides) for our modelling of mantle and core compositions. To first order, the planetary chemistry does not change, just the mineralogy. This practice is also in line with \citet{Dorn2015, Dorn2017, Dorn2017b}.
When we have the observational information of atmospheric thickness/composition of a terrestrial exoplanets at the era of JWST, we will know better the plausible pressure range associated with the solid regime of such a planet. With the further progress of ultrahigh pressure and temperature experiments of the physical state and properties of minerals, we can be more confident in our modelling of more complex exoplanetary mineralogies.

\subsection{Requirement on the precisions of host stellar abundances}

\citet{Hinkel2018} modelled the mineralogy for fictive planets around the 10 stars closest to the Sun and concluded that abundance uncertainties need to be on the order of [Fe/H] < 0.02 dex, [Si/H] < 0.01 dex, [Al/H] < 0.002 dex, [Mg/H] and [Ca/H] < 0.001 dex, in order to distinguish different planetary populations orbiting their sample of 10 stars. Based on our modelling results above, we have demonstrated that the interior composition and structure of those potential terrestrial exoplanets as studied can be distinguished from each other, by using the current, high-precision host stellar abundances (with a typical uncertainty of $\lesssim$ 0.04 dex). This is a much less strict requirement on the precisions of host stellar abundances. It is legitimate to strive for much more precise stellar abundances (as well as planetary mass and radius measurements) and thus to improve the interior modelling accuracy. However, we should also be realistic, since precisions of host stellar abundances depend not only on modelling techniques but also on characteristics of target stars and observation exposure time. A careful weighing of costs and benefits is crucial for the success of a space mission -- as noted in \cite{Dorn2015} -- like TESS/JWST and follow-up observations. Improving the exoplanetary interior modelling approaches may be an alternative and more sensible way to achieve better characterisation of planetary chemistry and structures in the coming decade. 

\subsection{Limitations}
\label{sec:limits}
The success of a theoretical model is subject to the limitations of premises/assumptions. Firstly, as mentioned earlier, our model is limited by how representative the Sun-to-Earth devolatilization pattern is for terrestrial planets in general. If we had a more exhaustive devolatilization model built on the comprehensive comparisons of the compositions of all inner solar system rocky bodies, gas giants, and icy bodies (including a variety of comets), we would have more accurate modelling results for exoplanetary chemistry. However, it is unclear to what extent the appearance of super-Earths in the proximity of their host stars (radically different to the scenario of the Solar System) would change the devolatilisation process and thus the bulk compositions of their planet companions in the circumstellar habitable zone.
We have just scratched the surface of considering devolatilisation in modelling exoplanetary chemistry and interiors, and this issue remains degenerate. Secondly, our approach to oxidise elements by following a sequence of oxidation might be worth revisiting if the oxidation of elements is done in parallel (simultaneously), not in sequence. In the parallel case, all rock-forming elements would compete to be oxidised to some extent. It is unclear how significant the end-member oxides resulting from this competition would be in comparison with those resulting from an assumed sequence. Thirdly, our modelling results are also limited to the current dimensionality of the model, namely, two-component overall structure (an iron alloy core and a silicate mantle), with no water layer (or oceans) or gas layer (or H/He envelopes) considered yet.
With the upcoming addition of planetary atmosphere observations, we are expecting to expand the dimensionality of our model to be inclusive of water/gas reservoirs in a terrestrial exoplanet by taking into account important atmospheric information, pressure and temperature associated with the actual size of the solid regime of a planetary body. 

Nevertheless, we envisage that if a terrestrial planet is confirmed in the habitable zone around the host stars studied here, our modelling results of the mantle and core compositions as well as core mass fractions listed in Table \ref{tab:comp} are good first-order estimates of the interior composition and structure of that planet. We also emphasize that the purpose of this work is 
to set more appropriate initial conditions for modelling exoplanetary interiors, especially with the addition of a \textquotedblleft devolatilisation\textquotedblright\thinspace process into the host stellar abundances. These enhanced constraints can be potentially transferable to some recent models \citep[e.g.][]{Dorn2017,Dorn2017b,Brugger2017,Unterborn2018} and thus enhance the model performance. Additional data (e.g., atmospheric information and applicable EoS parameters for all considered compositions) are required for further studies to decipher the detailed interior structure and chemistry, surface conditions and thus habitability of such terrestrial exoplanets. %

\section{Summary and Conclusions}
\label{sec:conclusion}
Devolatilisation (i.e. depletion of volatiles) plays an essential role in the formation of rocky planetary bodies from a stellar nebular disc \citep{Bland2005, Norris2017}. The terrestrial devolatilisation pattern of \cite{Wang2018b} has been applied in this study to infer the bulk elemental composition of potential terrestrial exoplanets that are presumed to exist within the circumstellar habitable zones. The inferred planetary bulk composition (rather than the host stellar abundances) provides improved principal constraints to model the interior composition and structure of such planets. Other recommended constraints include i) the mantle chemical network (SiO$_2$-CaO-Na$_2$O-MgO-Al$_2$O$_3$-FeO-NiO-SO$_3$, in order of the ease of oxidation); ii) the core chemical network (Fe-Ni-S alloy, Fe/Ni constrained at 18 $\pm$ 4 by number).

By applying these constraints to the Sun, we show that the mantle and core compositions of our model Earth are supported by the independent measurements/estimates for the planet Earth in the literature. By applying our modelling approach to selected planet hosts (Kepler-10, Kepler-20, Kepler-21 and Kepler-100), we find that the interior compositions and structures of potential terrestrial exoplanets in the habitable zones around these stars are diverse. For example, the estimates of core mass fraction range from about 1.5 wt\% (K10-exoE) to about 35 wt\% (K20-exoE). This diversity can be explained from the differences in their host stellar abundances.
With respect to the interior estimates (i.e. mantle and core compositions as well as core mass fraction), we conclude that a potential terrestrial planet orbiting Kepler-21 would be the most Earth-like while one orbiting Kepler-10 would be the least (among the cases we studied). High-precision host stellar abundances are critical for exoplanetary interior modelling. A precision better than $\sim$ 0.04 dex is necessary to assess the similarity of exoplanetary interiors to the Earth's, based on our modelling approach.

In summary, for more accurate estimates of interior composition and structure of terrestrial exoplanets, it is essential to enhance the initial conditions pertinent to planetary bulk composition and interior chemical models, in addition to the increasingly higher-precision measurements of host stellar photosphere and planetary mass and radius, alongside with the progressive observations of planetary atmospheres, in the era of TESS, PLATO and JWST. 

\section*{Acknowledgements}
We thank the anonymous referee whose comments greatly improved the quality of the paper. We acknowledge valuable discussions with William F. McDonough, Thomas Nordlander, and Stephen Mojzsis. 
HSW was supported by the Prime Minister's Australia Asia Endeavour Award (No. PMPGI-DCD-4014-2014) from Australian Government Department of Education and Training.  
FL was supported by the M$\"a$rta and Eric Holmberg Endowment from the Royal Physiographic Society of Lund. This work has made use of the Hypatia Catalog Database at \url{hypatiacatalog.com}, which was supported by NASA's Nexus for Exoplanet System Science (NExSS) research coordination network and the Vanderbilt Initiative in Data-Intensive Astrophysics (VIDA).

%%%%%%%%%%%%%%%%%%%%%%%%%%%%%%%%%%%%%%%%%%%%%%%%%%

%\clearpage
\bibliographystyle{mnras}
\bibliography{/Users/seanwhy/Documents/eLibrary/BibsTex/RockyExoModel} % 
%\bibliography{ref} % 
%%%%%%%%%%%%%%%%%%%%%%%%%%%%%%%%%%%%%%%%%%%%%%%%%%

%%%%%%%%%%%%%%%%% APPENDICES %%%%%%%%%%%%%%%%%%%%%
%\clearpage
\appendix

\section[Computational details of elemental fractionation between the mantle and the core]{Computational details of elemental fractionation between the mantle and the core}

\label{app:fractionation}
The elemental fractionation is based on the recommended chemical networks for the mantle and core of a terrestrial planet in Sects. \ref{sec:mantle} and \ref{sec:core}. The computational procedure illustrated in Figure \ref{fig:procedure} (with the source codes available on \url{https://github.com/astro-seanwhy/ExoInt}) is described in detail in the following. 

First, oxidise Si, Ca, Na, Mg, and Al (in order of the decreasing ease of oxidation addressed in Section \ref{sec:core}) to form oxides: SiO$_2$, CaO, Na$_2$O, MgO, and Al$_2$O$_3$, respectively. For each element, assess the sufficiency of O atoms: i) if the budget of (remaining) O atoms is not stoichiometrically sufficient to fully oxidise the element, then the non-oxidised part of the element's atoms and the atoms of the subsequent elements (up to Al) will be deemed as \textquotedblleft metals" in the mantle; ii) otherwise (namely, all atoms of an element can be fully oxidised), the oxidation process will be performed on the subsequent elements one by one (up to Al), unless the remaining O atoms have been exhausted.

Second, fractionate Fe, Ni, and S between the mantle and the core, for two scenarios: 

i) If the remaining O atoms (after having oxidised Si, Ca, Na, Mg, and Al) are still stoichiometrically sufficient to fully oxidise Fe, Ni, and S into the form of FeO, NiO, and SO$_3$, then there will be no Fe, Ni, or S fractionated into the core, and a coreless planet will form. The left over oxygen atoms may oxidise the stoichiometric amount of C atoms to form CO$_2$ in the phase of carbonates, then the remaining C atoms are present in the phase of graphite in the mantle. However, if the number of the left over oxygen atoms is more than that of C atoms, all C atoms will be oxidised to form CO$_2$ and the extra O is assumed to be present in the mantle to oxidise other minor elements that are not investigated in this work.

ii) Complementarily, if the remaining O atoms are \textit{not} stoichiometrically sufficient to fully oxidise Fe, Ni, and S, then the atoms of the three elements will be fractionated between the mantle and the core, and all C atoms will be present in the phase of graphite in the mantle. The upper limits of atomic abundances of Ni and S in the phase of NiO and SO$_3$ in the mantle are assumed to be equal to their respective planetary bulk abundances (after being devolatilised from the host stellar abundance). Their lower limits are assumed to be 0. Assume the abundances of Ni and S in the mantle can be any value within their respective upper and lower limits, following a uniform distribution (denoted as \textquoteleft $u$'): 
\begin{align}
N_{\textrm{Ni, mantle}} &=[0, N_{\textrm{Ni, bulk}}]_u \\
N_{\textrm{S, mantle}} &=[0, N_{\textrm{S, bulk}}]_u
\end{align}

The abundance of Fe in the phase of FeO (present in the mantle) can then be computed by 
\begin{equation}
N_{\textrm{Fe, mantle}}=N_{\textrm{O, remain}} - N_{\textrm{Ni, mantle}} - 3N_{\textrm{S, mantle}}
\end{equation}
where, $N_{\textrm{O, remain}}$ is the remaining amount of O atoms after Si, Ca, Na, Mg, and Al have been oxidised. 

Then, the corresponding abundances of Fe, Ni and S in the core can be obtained by deducting the abundance of them from the respective abundances of them in the bulk planet. Namely, 
\begin{align}
N_{\textrm{Ni, core}} &=N_{\textrm{Ni, bulk}} - N_{\textrm{Ni, mantle}} \\
N_{\textrm{S, core}} &=N_{\textrm{Fe, bulk}} - N_{\textrm{S, mantle}} \\
N_{\textrm{Fe, core}} &=N_{\textrm{Fe, bulk}} - N_{\textrm{Fe, mantle}}
\end{align}

On the basis, we use the constraint Fe/Ni = 18 $\pm$ 4 in the core (as addressed in Sect. \ref{sec:core}) to verify the series of Fe/Ni ratios as computed from each pair of Ni and S drawn from their respective uniform distributions (here, we draw $10^5$ times). Only computations matching this constraint are deemed as valid estimates. 

Then in combination with the atomic masses, these mantle and core abundances (by number) can be converted to the molar masses.
The core (molar) mass fraction is the total molar mass in the core divided by the sum of the total molar masses in both the mantle and the core.

The process above is iterated for each combination of planetary elemental abundances, by $2\times10^4$ Monte Carlo simulations assuming that each element's abundance (within its uncertainty) follows a Gaussian distribution. The results corresponding to the best fit of the pattern and the mean of each input elemental abundance are reported as the mean values of the modelling results (i.e. mantle composition, core composition, and core mass fraction) in Tables \ref{tab:verify} and \ref{tab:comp}. The error bar associated with a mean value corresponds to the standard deviation of the population of the modelling results for that quantity. The distribution of the modelling results for FeO, NiO, and SO$_3$ in the mantle and Fe, Ni, and S in the core as well as the core mass fraction may be asymmetric around their corresponding mean values. In this case, the standard deviation of the population of the modelling results for such a quantity is not an accurate but conservative estimate for the uncertainty on the mean value of the quantity. If the standard deviation may cause the lower limit of the quantity to be negative, then the lower limit is set to be zero. As such, the error bars of some quantities in Tables \ref{tab:verify} and \ref{tab:comp} are asymmetric.

%%%%%%%%%%%%%%%%%%%%%%%%%%%%%%%%%%%%%%%%%%%%%%%%%%%%%%%%%%%%%%%%%%%%%%%%%%%%%%%

%%%%%%%%%%%%%%%%%%%%%%%%%%%%%%%%%%%%%%%%%%%%%%%%%%

% Don't change these lines
\bsp	% typesetting comment
\label{lastpage}
\end{document}